\documentclass[12pt,letterpaper]{article}
\pdfoutput=1
\usepackage{wrapfig}
\usepackage{subfigure}
\usepackage{graphicx}
\usepackage{amsfonts}
\usepackage{color}
\makeatletter \@addtoreset{equation}{section} \makeatother

\begin{document}

\begin{titlepage}

\begin{center}
{\Large \bf Chaos in Lifshitz Spacetimes} \vskip .7 cm \vskip 1 cm
{\large   Xiaojian Bai${}^{1}$, Junde Chen${}^{2}$, Bum-Hoon
Lee${}^{1}$, and Taeyoon Moon${}^3$}
\end{center}
\vskip .4cm \centerline{\it ${}^1$Center for Quantum Spacetime,
Sogang University, Seoul, Korea} \vskip .4cm \centerline{\it
${}^2$Department of Physics, Shandong University, Jinan,
China}\vskip .4cm \centerline{\it ${}^3$Institute of Basic Sciences
and School of Computer Aided Science,} \centerline{\it Inje
University, Gimhae, Korea} \vskip .4cm {\tt   baixj@sogang.ac.kr,
chunte@mail.sdu.edu.cn, bhl@sogang.ac.kr, tymoon@inje.ac.kr} \vskip
3cm

\begin{abstract}
We investigate the chaotic behavior of a circular test string in the
Lifshitz spacetimes considering the critical exponent $z$ as an
external control parameter. It is demonstrated that two primary
tools to observe chaos in this system are Poincar\'{e} section and
Lyapunov exponent. Finally, the numerical result shows that if
$z=1$, the string dynamics is regular, while in a case slightly
larger than $z=1$, the dynamics can be irregular and chaotic.
\end{abstract}
\end{titlepage}

\setcounter{page}{1} \renewcommand{\thefootnote}{\arabic{footnote}}
\setcounter{footnote}{0}

\section{Introduction}
The AdS/CFT correspondence
\cite{Maldacena:1997re,Gubser:1998bc,Witten:1998qj} is a powerful
tool to describe strongly correlated system in terms of weakly
coupled gravitational dynamics, and vice versa. Recently, it has
been conjectured, on the analogy of the relativistic AdS/CFT
correspondence, the existence of a similar holographic dual
description in condensed matter physics
\cite{Hartnoll:2009sz,McGreevy:2009xe}, which corresponds to
non-relativistic system without Lorentz invariance. In such a
theory, scaling transformation of time and space take the form
anisotropically
\begin{eqnarray}\label{tx}
t\to\lambda^z t,~~~~~x\to\lambda x,
\end{eqnarray}
where $z$ is a dynamical critical exponent, revealing the anisotropy
between space and time whenever $z\neq1$. It is well-known that the
so called {\em Lifshtiz metric} with scaling symmetry (\ref{tx}) in
$D$ dimensions provides a geometric description of Lifshitz
spacetimes
\cite{Koroteev:2007yp,Balasubramanian:2008dm,Kachru:2008yh}:
\begin{eqnarray}
ds^2=\ell^2\left(-r^{2z}dt^2+\frac{dr^2}{r^2}+r^2dx_i^2\right),
\end{eqnarray}
where $i=2,3,\cdots, D-1$. However, it was shown in the literature
\cite{Hartnoll:2009sz,Kachru:2008yh,Copsey:2010ya} that the Lifshitz
geometry ($z\neq1$) has a problem of null curvature singularity by
calculating the tidal forces between infalling geodesics. More
precisely, in Ref. \cite{Horowitz:2011gh} the authors demonstrated
that Lifshitz solutions are unstable by showing that the test
strings become infinitely excited when crossing the singularity.
Very recently, it was suggested in Ref.\cite{Bao:2012yt} that
Lifshitz singularity problem can be resolved by recognizing that
string propagation is modified enough to avoid singular tidal forces
when considering nontrivial matter contents, which corresponds to
stress-energy sources.

 In this paper,
we investigate this kind of the Lifshitz instability which is
present whenever $z\neq1$, from an alternative point of view, i.e.,
{\it chaos}. It is demonstrated that dynamical instability of a
nonlinear system, which is described by a nonlinearity of the
equations of motion in most cases, could be provided chaoticity of
the system. It is well known that a test particle motion is
completely regular, that is, {\it integrable} ({\it non-chaotic}) in
the generic Kerr-Newmann background \cite{carter} and the particle
geodesic equation can be analytically solved in higher dimensional
spherically symmetric spacetimes \cite{Hackmann:2008tu}. On the
contrary of the case of the particle geodesic, it was shown that the
motion of test strings can exhibit chaotic behavior in spherically
symmetric black hole \cite{Frolov:1999pj}, AdS soliton
\cite{Basu:2011dg}, and AdS$_{5}\times$ T$^{1,1}$ \cite{Basu:2011di}
background.  More interestingly, in Ref. \cite{Zayas:2010fs} the
chaotic behavior of the classical string was studied in the context
of the gauge/gravity correspondence.

Along this line of research containing the test string, we intend to
investigate the chaotic behavior of the circular test string moving
in the Lifshitz spacetimes by thinking of the critical exponent $z$
as an external control parameter. A key motivation for introducing
the control parameter $z$ is to examine the relation between space
time anisotropy and chaos. Our quantitative approaches to
identifying and measuring chaos in this system include two numerical
techniques; one is to check for the breaking of the KAM
(Kolmogorov-Arnold-Moser) tori \cite{kam} through the numerical
computation of the Poincar\'{e} sections \cite{poincare} and the
other is to calculate the maximum Lyapunov exponent $\lambda$, in
which the defining signature of chaos takes the positive $\lambda$.
Using both the poincar\'{e} sections and Lyapunov exponent in the
Lifshitz spacetimes, we show explicitly that if $z=1$, the string
dynamics is regular, while in the case slightly off $z=1$, the
dynamics can be irregular and chaotic.

\section{Instability of the Lifshitz spacetimes}

 Let us begin by
introducing the following Lifshitz metric\footnote{Here, we consider
a spherically symmetric form of the metric for convenience to
describe a circular test string. Note that the above metric can be
obtained from choosing $d=3,~k=1,~m=\rho_j=0$ in Eq. (2.24) of the
literature \cite{Tarrio:2011de}.} \cite{Tarrio:2011de}:
\begin{eqnarray}\label{metric}
ds^2=-\frac{r^{2z-2}}{\ell^{2
z-2}}f(r)dt^2+\frac{1}{f(r)}dr^2+r^2d\theta^2+r^2\sin^2\theta
d\varphi^2,
\end{eqnarray}
where $z$ denotes the critical exponent, $\ell$ is the curvature
radius of Lifshitz spacetimes, and metric function $f(r)$ is given
by
\begin{eqnarray}
f(r)=\frac{1}{z^2}+\frac{r^2}{\ell^2}.
\end{eqnarray}
We note that the Lifshitz metric (\ref{metric}) reduces simply to
AdS spacetimes when $z=1$. However, if $z\neq1$, as was shown in
\cite{Horowitz:2011gh}, the tidal forces in the Lifshitz geometry
diverge as $(z-1)/r^{2z}$, which implies that there is a curvature
singularity at $r=0$. To see this more concretely, we consider a
radial timelike geodesic with the normalization $U_{\mu}U^{\mu}=-1$,
where tangent vector $U^{\mu}$ is given by
$U^{\mu}=(\dot{t},\dot{r},0,0)$. Here the dot `$\cdot$' denotes the
differentiation with respect to the proper time $\tau$. Now we
introduce a parallel orthonormal frame field along the geodesic
\begin{eqnarray}
e_{(0)}^{~\mu}&=&\frac{E}{fr^{2(z-1)}}\left(\frac{\partial}{\partial
t}\right)^{\mu}-\frac{E}{r^{z-1}}
\sqrt{\frac{1}{\ell^{2z-2}}-\frac{fr^{2(z-1)}}{E^2}}\left(\frac{\partial}{\partial
r}\right)^{\mu},\nonumber\\
&&\nonumber\\
e_{(1)}^{~\mu}&=&\frac{E}{fr^{2(z-1)}}
\sqrt{1-\frac{fr^{2(z-1)}\ell^{2(z-1)}}{E^2}}\left(\frac{\partial}{\partial
t}\right)^{\mu}-\frac{E}{r^{z-1}\ell^{z-1}}\left(\frac{\partial}{\partial
r}\right)^{\mu},\nonumber\\
&&\nonumber\\
e_{(2)}^{~\mu}&=&\frac{1}{r}\left(\frac{\partial}{\partial\theta}\right)^{\mu}
,~~~~e_{(3)}^{~\mu}~=~\frac{1}{r\sin\theta}\left(\frac{\partial}{\partial\varphi}\right)^{\mu},
\nonumber
\end{eqnarray}
where Greek indices denote spacetime coordinates
$(t,r,\theta,\varphi)$ and indices enclosed in parentheses describe
the flat tangent space in which the orthonormal frame is defined. In
the above expressions, the conserved energy $E$ can be written by
$E=\dot{t}fr^{2(z-1)}$. It turns out that in this orthonormal frame,
we have the non-vanishing components of the Riemann tensor as
follows:
\begin{eqnarray}
&&\hspace*{-3.5em}R_{(0)(1)(0)(1)}=\frac{(z-1)(z-2)}{z^2r^2}+\frac{z^2}{\ell^2},
~~~~R_{(0)(2)(0)(2)}=R_{(0)(3)(0)(3)}=\frac{E^2\ell^{2(1-z)}(z-1)}{r^{2z}}+\frac{1}{\ell^2}\\
&&\nonumber\\
&&\hspace*{-3.5em}R_{(2)(3)(2)(3)}=\frac{z^2-1}{z^2r^2}-\frac{1}{\ell^2},~~~~R_{(1)(2)(1)(2)}=R_{(1)(3)(1)(3)}=\frac{E^2\ell^{2(1-z)}(z-1)}{r^{2z}}
-\frac{z-1}{z^2r^2}-\frac{z}{\ell^2}\\
&&\nonumber\\
&&\hspace*{-3.5em}R_{(0)(2)(1)(2)}=R_{(0)(3)(1)(3)}=\frac{E^2\ell^{2(1-z)}(z-1)}{r^{2z}}
\sqrt{1-\frac{\ell^{2z-4}r^{2(z-1)}(\ell^2+z^2r^2)}{E^2z^2}},
\end{eqnarray}
where the Riemann tensor in the orthonormal frame is defined by
$R_{(a)(b)(c)(d)}=R_{\mu\nu\rho\sigma}e_{(a)}^{~\mu}
e_{(b)}^{~\nu}e_{(c)}^{~\rho}e_{(d)}^{~\sigma}$. The resulting
components of the Riemann tensor show that whenever $z\neq1$, the
Lifshitz geometry is unstable due to the divergence as
$(z-1)/r^{2z}$ or $(z-1)/r^{2}$ in the limit $r\to0$. Most of all,
it is demonstrated in the literature \cite{Horowitz:2011gh} that
Lifshitz solutions can be unstable\footnote{The authors in Ref.
\cite{Bao:2012yt} presented that this kind of pathological behavior
would be resolved by noticing that the string propagation can be
modified enough to avoid singular tidal forces when taking into
account nontrivial matter contents, see also
\cite{Harrison:2012vy}.} by showing that the test strings become
infinitely excited when crossing the singularity. In the present
paper, however, we study the nature of this instability in an
alternative point of view, i.e., chaos. It is of importance to note
that the presence of chaos in a given spacetimes would be a
geometrical criterion of local instability. On the other hand, the
chaotic behavior of a test object moving in the system can be a key
observation of instability when examining if the system is stable or
not. For this purpose, we investigate the chaotic behavior of a
circular test string moving near $r=0$ in the Lifshitz spacetimes by
considering the critical exponent $z$ as an external control
parameter. Before proceeding further, we first introduce a circular
string moving in the Lifshitz spacetimes in the next section.

\section{Circular string in the Lifshitz spacetimes}

 In order to investigate the behavior of a test string in the Lifshitz spacetimes,
 let us first consider the Nambu-Goto action
\begin{eqnarray}\label{pol}
S=-T\int~d\tau d\sigma\sqrt{-{\rm det}[h_{ab}]} ~~{\rm
with}~~h_{ab}=g_{\mu\nu}\partial_aX^{\mu}\partial_bX^{\nu},
\end{eqnarray}
where $T$ is the string tension, $h_{ab}$ is the induced metric on
the worldsheet, $g_{\mu\nu}$ is the spacetime metric, and
$X^{\mu}=X^{\mu}(\tau,\sigma)$ are the spacetime coordinates of the
string. The index $a$ and $b$ in the action (\ref{pol}) run over
values $(\tau,\sigma)$.
 Varying the action (\ref{pol}) with respect to $X^{\mu}$ leads to the string equations
 of motion
\begin{eqnarray}
\ddot{X}^{\mu}-X^{\prime\prime\mu}+
\Gamma^{\mu}_{\rho\sigma}(\dot{X}^{\rho}\dot{X}^{\sigma}-X^{\prime\rho}X^{\prime\sigma})=0,
\label{meq}
\end{eqnarray}
where the dot `$\cdot$' and prime `$\prime$' denote the
differentiation with respect to the string world sheet coordinates
$\tau$ and $\sigma$, respectively. On the other hand, Weyl
invariance and reparametrization invariance allow us to choose the
conformal gauge $h_{ab}=\Omega^2\eta_{ab}$, where $\eta_{ab}$ is
locally flat and $\Omega$ is a scalar function of the worldsheet
coordinates $(\tau,\sigma)$. It turns out that in this gauge, the
corresponding conditions are given by $h_{\tau\sigma}=0$ and
$h_{\tau\tau}+h_{\sigma\sigma}=0$, which lead to the constraints
\begin{eqnarray}
g_{\mu\nu}\dot{X}^{\mu}X^{\prime\nu}
=g_{\mu\nu}(\dot{X}^{\mu}\dot{X}^{\nu}+X^{\prime\mu}X^{\prime\nu})=0.
\label{conts}
\end{eqnarray}
 Now we consider the circular string, which is obtained by the ansatz:
\begin{eqnarray}\label{ant}
t=t(\tau),~~~r=r(\tau),~~~\theta=\theta(\tau),~~~\varphi=\sigma.
\end{eqnarray}
With the circular string ansatz (\ref{ant}), the equations of motion
(\ref{meq}) and constraints (\ref{conts}) in the Lifshitz background
(\ref{metric}) yield
\begin{eqnarray}
&&\dot{t}~=~\frac{E}{fr^{2(z-1)}}\nonumber\\
&&\nonumber\\ &&\ddot{r}~=~\frac{1-z}{r}\dot{r}^2 +\Big(
(2-z)rf-\frac{r^2}{2}f^{\prime}\Big)\dot{\theta}^2-(zrf+\frac{r^2}{2}f^{\prime})\sin^2\theta=0
\label{ieq1}\\
&&\nonumber\\
&&\ddot{\theta}=-\frac{2}{r}\dot{r}\dot{\theta}-\sin\theta\cos\theta\label{ieq2}\\
&&\nonumber\\
&&\ell^{2z-2}r^{2z-2}\Big(\dot{r}^2+fr^2(\dot{\theta}^2+\sin^2\theta)\Big)=E^2\label{con}
\end{eqnarray}
Note that the constraint (\ref{con}) is the same with the one
obtained from the Hamiltonian constraint $H=0$ for the following
Hamiltonian
\begin{eqnarray}
H=\frac{1}{2}fP_{r}^2+\frac{1}{2r^2}P_{\theta}^2-\frac{\ell^{2-2z}
r^{2-2z}}{2f}E^2+\frac{1}{2}r^2\sin^2\theta,
\end{eqnarray}
where $P_{r}=\dot{r}/f$ and $P_{\theta}=r^2\dot\theta$.

We mention that it is a formidable task to solve the above equations
(\ref{ieq1})$\sim$(\ref{con}) analytically, because they are
non-integrable. Thus we will analyze the equations numerically in
the next section. We remind that our primary purpose is to
investigate the chaoticity of a given non-integrable system in such
a way that we treat the critical exponent $z$ as an external control
parameter and we will use a couple of quantitative approaches to
identifying and measuring chaos; Poincar\'{e} sections and Lyapunov
exponent.

\section{Numerical results}
\subsection*{Poincar\'{e} sections}
\begin{figure}[tbph]
\begin{center}
\begin{tabular}{cc}
\includegraphics[width=.32\linewidth,origin=tl]{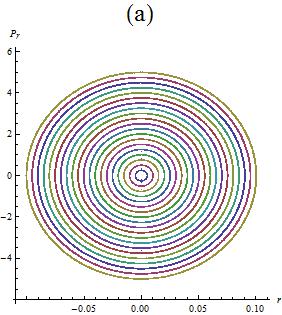}
, &\quad\quad\includegraphics[width=.32\linewidth,origin=tl]{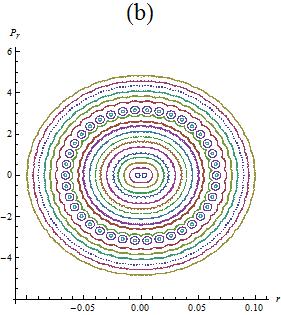}\\
\includegraphics[width=.32\linewidth,origin=tl]{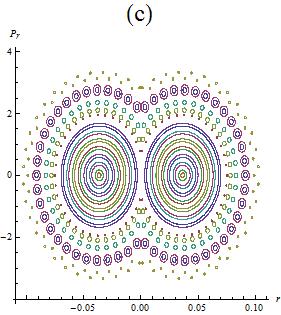}
, &\quad\quad\includegraphics[width=.32\linewidth,origin=tl]{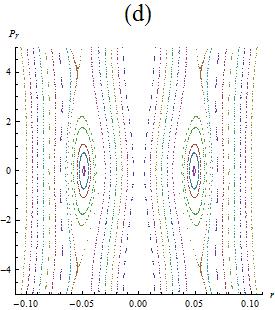}\\
\includegraphics[width=.32\linewidth,origin=tl]{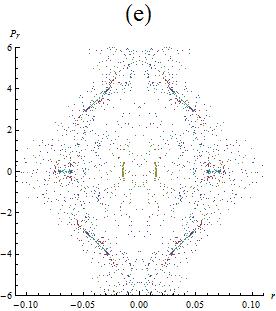}
,
&\quad\quad\includegraphics[width=.32\linewidth,origin=tl]{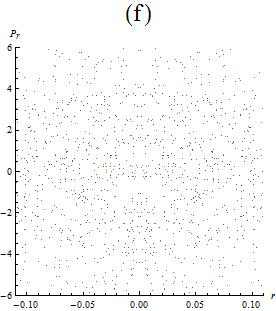}
\end{tabular}
\end{center}
\caption{The Poincar\'{e} sections for $E=50$, $\ell=1$ in {\it
isotropic} (a) and {\it anisotropic} (b)$\sim$(f) cases. Fig. 1(a)
corresponds to $z=1$ and (b) to 1.00008, (c) 1.001, (d) 1.025, (e)
1.03, (f) 1.045. The figure 1(a) consists of the sets of KAM tori
only ({\it non chaotic}), while for the figures (b)$\sim$(e) the KAM
tori break up and it is eventually shown the Cantor sets in Fig.
1(f).}\label{fig:phi1}
\end{figure}
In order to investigate whether the string's particular orbit can
show the chaotic behavior or not, we first check for the breaking of
the KAM tori which is indicative of chaos for a given system by
numerical computation of the Poincar\'{e} sections in the phase
space ($r,P_r$). On the other hand, if Poincar\'{e} sections consist
of continuous closed lines without revealing an essentially random
series of dots, the system is not chaotic. It is of importance to
note that since  our main interest is to explore the connection
between chaotic behavior and spacetime anisotropy ($z\neq1$), the
critical exponent $z$ in our numerical analysis plays a central role
of an external control parameter. To this end, we shall focus on the
cases $z=1$ (isotropic) and slightly larger (anisotropic) than
$z=1$. Note that for fixed $E$ and $\ell$, we set initial values of
($\dot{r},\theta$) at $\tau=0$ as $\dot{r}(0)=0$ and $\theta(0)=0$.
The initial value $\dot{\theta}(0)$ can be determined by the
constraint equation (\ref{con}) when varying a value of $r(0)$. In
other words, the string trajectories for various initial conditions
can be obtained from integrating equations (\ref{ieq1}) and
(\ref{ieq2}). As a result, the Poincar\'{e} sections for different
initial conditions and fixed $E=50,~\ell=1$ are presented in Fig.
1(a) and (b)$\sim$ (f), illustrated with $z=1$ and $z\neq1$,
respectively. In these figures different colors correspond to
different values of $r(0)$.

A few explanations for the figure $1$ are in order. First, Fig. 1(a)
consists of KAM tori only, which implies that the corresponding
motion is non chaotic. Whereas, Fig. 1(b) being slightly off $z=1$
shows that some KAM tori break up into discrete segments and new
sets of smaller tori emerge. Second, being a few more off the point
$z=1$ (see Fig. 1(c)), it seems that the largest KAM torus in Fig.
1(b) is decomposed into two smaller tori and the total number of
tori is much more than that in Fig. 1(b). The figure 1(d) indicates
that most of the tori break up and eventually scatter. In the last,
there exist the sets of sparse points in almost whole (Fig. 1(e))
and full ranges (Fig. 1(f)) corresponding to the Cantor sets ({\it
cantori}).

\newpage
\subsection*{Lyapunov exponent}

A more quantitative method one uses to investigate chaos is to
calculate the largest Lyapunov exponent in a given system. If we
consider the growth rate between two initially nearby trajectories,
the largest Lyapunov exponent $\lambda$ is defined as
\begin{eqnarray}\label{lya}
\lambda(t)=\frac{1}{t}\ln\Big[\frac{dX(t)}{dX(0)}\Big],
\end{eqnarray}
where $dX$ is the difference between two points in phase space,
which corresponds, more precisely, to the Cartesian distance between
$(r,\dot{r},\theta,\dot{\theta})$ and $(\tilde{r},{\dot{\tilde{r}}},
\tilde\theta,{\dot{\tilde\theta}})$ of two nearby phase space
trajectories. We note that in a chaotic system, the growth rate in
logarithm in Eq.(\ref{lya}) will be exponential in time with a
positive constant $\lambda$, while for a regular system the rate is
expected to be linear or power law growth in time, which implies
that the Largest Lyapunov exponent $\lambda$ goes to zero in the
limit $t\to\infty$. Thus, it is obvious that our goal in this
section is to observe the positive Lyapunov exponent, which may be
taken as the defining signature of chaos.

On the other hand, our numerical analysis begins with a set of
initial values for $z=1.01$, $E=20,~\ell=1,$ which is given as
follows: $r(0)=1/2,~\dot{r}(0)=0,~\theta(0)=0$, and
$~\dot{\theta}(0)=36.31,$ where $\dot{\theta}(0)$ is determined by
the constraint equation (\ref{con}). We also consider an initial
tiny separation as $\delta_0=10^{-3}$ for a nearby point
$r(0)+\delta_0$. With these initial values, we try to compute the
Lyapunov exponent [Eq.(\ref{lya})] following the algorithm provided
by Sprott \cite{jcsprott,sprott1}. As a consequence, we present in
figure 2 the result of estimating the largest Lyapunov exponent with
a convergent positive value, which implies that the motion can be
chaotic when $z\neq1$. On the contrary, it is checked that the
Lyapunov exponent becomes almost zero ($\sim10^{-3}$) for $z=1$,
which corresponds to non-chaotic motion.
\begin{figure*}[t!]
   \centering
   \includegraphics{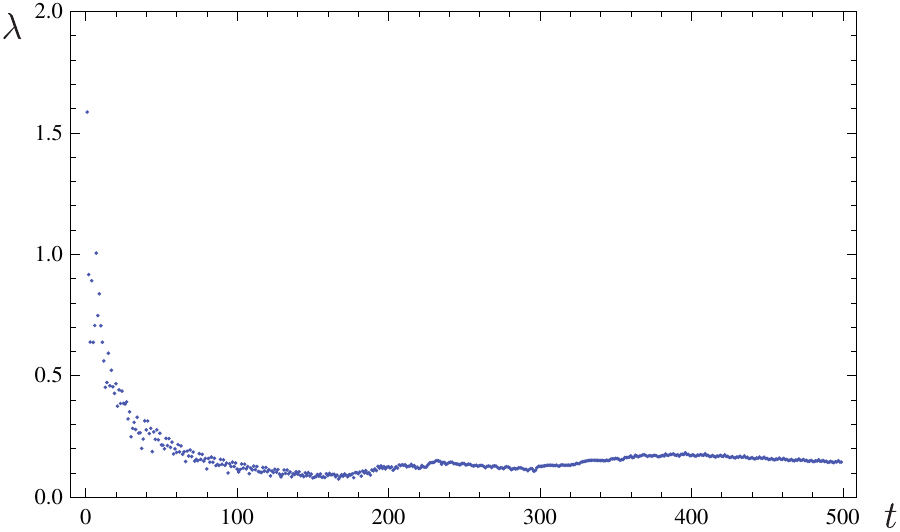}
\caption{The Lyapunov exponent $\lambda$ graph as function of time
$t$ for $z=1.01$, $E=20,$ and $\ell=1$. In this figure, the Lyapunov
exponent converges asymptotically to one positive value of $0.15$.}
\end{figure*}
\newpage
\section{Conclusion}

In this work, we investigated the chaotic behavior of the circular
test string moving in the Lifshitz background, which allows the
scaling anisotropy between space and time whenever $z\neq1$. Most of
all, we considered that the critical exponent $z$ defined in the
Lifshitz background plays a role of the external control parameter,
since our primary goal was to examine the relation between space
time anisotropy and chaos. This is clearly different from the case
of the literatures
\cite{Frolov:1999pj,Basu:2011dg,Basu:2011di,Zayas:2010fs}, which
were thought of the total conserved energy of the string as the
external control parameter to investigate the chaotic behavior of
the test string. Our numerical results obtained by using both the
Poincar\'{e} section and Lyapunov exponent indicate that if $z=1$,
the motion of the string is regular, while in the case slightly off
$z=1$, its behavior can be irregular and chaotic.

Finally we propose that the space time anisotropy which breaks
Lorentz symmetry may cause the system to be chaotic. However, in
order to generalize the present result beyond the Lifshitz
background that we considered in this paper, we need to explore the
chaoticity of the system, which is present in the Lifshitz
spacetimes for arbitrary $z$ with black hole configurations
\cite{Tarrio:2011de,lif} or naked singularity \cite{Tarrio:2011de},
which will be studied elsewhere.

We conclude a remark on the issue related to extending the
Lifshitz/CFT correspondence. The authors in the literature
\cite{Zayas:2010fs} suggested the extension of AdS/CFT
correspondence to include chaotic dynamics in such a way that the
proper bound for the Poincar\'{e} recurrence time could be connected
with the positive largest Lyapunov exponent, which describes the
late time behavior of the distance between two string worldsheets as
the correlator to which the worldsheets are dual. The explicit
computation of the proper bound for the Poincar\'{e} recurrence time
in the Lifshitz background is beyond the scope of the present paper,
but nevertheless, it is worthwhile to be explored in future work.

\subsection*{Acknowledgements}
~~~~~T.M. would like to thank Prof. Yun Soo Myung for his valuable
comments. This work was supported by the National Research
Foundation of Korea(NRF) grant funded with grant number
2014R1A2A1A01002306. T.M. was supported by the National Research
Foundation of Korea (NRF) grant funded by the Korea government
(MEST) (No.2012-R1A1A2A10040499).

\end{document}